\documentstyle[12pt]{article}

\begin{document}

\begin{center}
{\large {\bf SELF-SIMILAR EXTRAPOLATION FOR THE LAW OF
ACOUSTIC EMISSION BEFORE FAILURE OF HETEROGENEOUS MATERIALS} \\ [3mm]

A. Moura$^1$ and V.I. Yukalov$^2$} \\ [2mm]
{\it $^1$GEMPPM, UMR CNRS 5510 (Bat. B. Pascal)\\
Institut National des Sciences Appliquees de Lyon \\
20, av. A. Einstein, 69 621 Villeurbanne, France \\
e-mail: andre.moura@insa-lyon.fr \\ [3mm]

$^2$Bogolubov Laboratory of Theoretical Physics \\
Joint Institute for Nuclear Research, Dubna 141980, Russia \\
e-mail: yukalov@thsun1.jinr.ru}
\end{center}

\vskip 2cm

\begin{abstract}

Acoustic emission before the failure of heterogeneous materials
is studied as a function of applied hydrostatic pressure. A formula for
the energy release is suggested, which is valid in the whole diapason of
pressures, from zero to the critical pressure of rupture. This formula is
obtained by employing the extrapolation technique of the self-similar
approximation theory. The result is fitted to experiment in order to
demonstrate the correct general behaviour of the obtained expression for
the energy release.
\end{abstract}

\section{Introduction}

There exists a widespread understanding in scientific community that the
moment of rupture in heterogeneous materials is somewhat similar to a
critical point. The fracturing process is then a kind of a second-order
phase transition [2,4]. This underlies the intuitive assumption which compares
global failure in disordered materials at mesoscopic scale with percolation
at microscopic scale. So, as percolation, global failure is assumed to be
a critical phenomenon. The fracturing process is then classified as a
self-organised random irreversible process. Extensive numerical simulations
and experimental measurements have confirmed this conclusion. This concerns
the global failure of materials as well as earthquakes [9], since an
earthquake can also be considered as a global failure of a large material
mass. The classical power law, defining the energy release rate, has been
verified in the critical region, close to the time of either a large
earthquake or a global materials failure. The related critical exponent
is, in general, a complex number, which is due to taking into account
long-range elastic interactions transported by stress fields around defects
and cracks. The imaginary part of the exponent gives rise to the so-called
log-periodic corrections which have been identified quite early in
renormalization group solutions for critical phase transitions [7,10].
Experimental results of acoustic emission measurements and seismograms
unambiguously revealed the existence of oscillatory corrections before
materials fractures and earthquakes, respectively [1,10]. If one would
know the general law describing this acoustic emission, one would be able
to make an early prediction of the global failure in heterogeneous materials.
It is a principal, not yet solved problem, to find such a general law,
which would be valid not solely in the asymptotic vicinity of the critical
point, but in the whole interval of pressures or times, both near the critical
point of failure as well as far from it. And this is the main aim of the
present paper to suggest a mathematically grounded derivation for such
a law and to confront it to available experimental data [5]. For this purpose,
we employ the recent progress in the self-similar approximation theory [12--18]
which has been successfully applied to the theory of critical phenomena and
to time series forecasting [19--21]. We would like to stress here that this
theory is general and can be used for any given series of experimental data,
whether this concerns phase transitions, financial markets, or any other
series.

\section{Modelling}

Let us be interested in the behaviour of the energy release $E(p)$
as a function of pressure $p$, considered between $p\approx 0$, i.e. at the
initial stage when the very early damages occur, and up to the point $p=p_c$
of global failure. Note that $p$ refers to fracturing under applied spatially
uniform loading, as in experiments where water pressure inside a spherical tank
increases up to the material failure. Experimentally, the applied pressure is
usually increased linearly with time. The behaviour of $E(p)$ at the vicinity
of the critical point of rupture is
\begin{equation}
\label{1}
E(p) \approx E(p_c) + A(p_c - p)^\alpha + B(p_c-p)^\alpha\cos[\omega\ln
(p_c-p)+\varphi] \; ,
\end{equation}
as $p\rightarrow p_c$, where $A,B,\alpha,\omega$ and $\varphi$ are parameters
($\alpha$ and $\omega$ are the real and imaginary parts, respectively,
of the so-called critical complex exponent). The problem is how to find the
behaviour of $E(p)$ in the whole range $0\leq p \leq p_c$? Here we consider
the case of one variable, pressure. A more general loading, involving not one
variable, like $p$, but several variables can also be treated by the theory.
However, before being involved in such generalisations, we would like to show
on a simpler  example how to accomplish the extrapolation. Our main aim in
this letter is to demonstrate the general possibility of realising such an
accurate extrapolation for the energy release.

Let us transform Eq. (1) by introducing the dimensionless variable
$x\equiv(p_c-p)/p_c$, $(0\leq x\leq 1)$ and the reduced energy release
$f(x)\equiv E(p)/E(p_c)$  which is a dimensionless function. Then, expansion
(1) takes the form
\begin{equation}
\label{2}
f(x) \approx 1 + \tilde a(x)\; x^\alpha \; , \qquad
\tilde a(x) \equiv \lambda[1+\mu\cos(\omega\ln x +\beta)]
\end{equation}
whose parameters $\beta$, $\lambda$, and $\mu$ can be easily expressed through
$p_c$, $E(p_c)$, $A$, $B$, $\omega$, and $\varphi$. Note that function
$\tilde a(x)$ cannot be expanded in powers of $x$ near zero. Therefore,
expansion (2) can be treated as a generalised asymptotic series, as defined
by Poincare [8], with $\tilde a(x)$ considered as a coefficient. Employing
the self-similar approximation theory [12--21], the asymptotic series (2)
can be extrapolated to
\begin{equation}
\label{3}
f^*(x)=\exp[c(x)\; x^\alpha] \; , \qquad c(x) = [1 +\mu\cos(\omega\ln x +
\beta)]\; \tau \; ,
\end{equation}
where $\tau$ has to be defined from an optimization or boundary condition.
As is clear, the boundary condition $E(0)\approx 0$  is not convenient here,
leading to large errors in defining $\tau$ because of the physical impossibility
to determine the very early precursors with a high accuracy. A more judicious
choice is to consider a global physical quantity. In the same way as treating
$E(p)$ as the integrated energy release rate, we may define $F(p)$ as the integral
energy release $\int_0^p E(q)dq$. Given $p_0$, the parameter $\tau$ then can be
defined from the sum optimization rule
\begin{equation}
\label{4}
\int_{y_0=\frac{p_c-p_0}{p_c}}^1 \; f^*(x)\; dx = \frac{F(p_0)}{p_cE(p_c)} \qquad
(y_0\rightarrow 0, \;  p_0\rightarrow p_c-0) \; .
\end{equation}
In general, the lower limit in the integral (4) can pertain to the whole interval
[0,1]. In reality, the trustful experimental data are available only below $p_c$.
One has to choose the data at the highest available pressure, so that the lower
limit in the sum rule (4) be sufficiently small.

In this way, the self-similar approximant (3) becomes
\begin{equation}
\label{5}
f^*(x) =\exp\left\{ \left [ 1 +\mu\cos(\omega\ln x +\beta)\right ] \;
\tau x^\alpha \right \}
\end{equation}
with $\tau$ given by the sum rule (4). Note that this formula contains the same
number of parameters as the initial formula (1). Returning, with the help
of  relations (1) and (2), to the energy release,  we come to the formula
\begin{equation}
\label{6}
E^*(p) = E(p_c)\; f^*(x) \; ,
\end{equation}
which extrapolates $E(p)$ for the whole interval $0\leq p\leq p_c$. It is worth
stressing that the law (6) is obtained as a self-similar extrapolation of the
acoustic emission signals observed at the vicinity of the rupture. Another
possibility would be to base this kind of extrapolation, starting from the signals
existing at the initial stage of the process, being yet very far from the rupture.
This latter way was used by Gluzman et al. [3], who started with a given
polynomial expansion at the early time of acoustic emissions, treating the power
law behaviour close to rupture as a boundary condition. To our mind, this opposite
approach has two weak points: First of all, the early acoustical precursors, as
is known, are very small, being embodied in the acoustical noise, because of
which they can be hardly measured with a good accuracy. Second, it looks
difficult or even impossible to extract information on log-periodic corrections
from the early acoustic signals. Another phenomenological expression for the
energy release in the noncritical region has been suggested by Sornette and
Andersen [10] whose arguments were based on the existence of a scaling of the
macroscopic elastic modulus and the elastic energy release rate in the thermal
fuse model [6]. The following phenomenological form has been proposed [5]
\begin{equation}
\label{7}
E(p) \approx E(p_c) A\left ( {\rm tanh}\;\frac{p_c-p}{\tau_0}\right )^\alpha +
B\left ( {\rm tanh}\;\frac{p_c-p}{\tau_0}\right )^\alpha\;
\cos\left\{ \omega\left ( {\rm tanh}\;\frac{p_c-p}{\tau_o}\right ) +\varphi
\right \} \; ,
\end{equation}
which is a pure power law, like Eq. (1), in the critical region close to
rupture and exponentially relaxes in the noncritical region far from rupture,
where only a few damages occur. However, the suggested form (7) possesses an
unphysical behaviour, being negative in a wide range of its variable. At the
same time, our self-similar formula (5) yields the energy release (6) that
is always positive. In Fig. 1, we compare the behaviour of $E(p)$ , in the
dimensionless form, for both Eqs. (6) and (7), with the parameters chosen
so that to have similar variations when approaching the critical point. In
this figure, one can clearly see that Eq. (7) possesses unphysical negative
values for the major part of the interval [0,1], while Eq. (6) is everywhere
positive.

\section{Comparison with experimental data}

In order to fit our self-similar formula to experimental results of acoustic
emission measurements, we refer to the most accurate data, available nowadays,
that is those of Anifrani's team [1] at Aerospatiale-Matra Inc., which are
reported by Johansen and Sornette [5]. In these measurements, the acoustic
emission has been recorded during the pressurisation of spherical tanks of
kevlar or carbon fibres pre-impregnated in a resin matrix wrapped up around
a thin metallic liner (steel or titanium) fabricated and instrumented by
Aerospatiale-Matra Inc. It has been found that the seven acoustic emission
recordings of seven pressure tanks, which were brought to rupture, exhibit
a clear acceleration in agreement with a power law divergence as expected by
the critical point theory. A strong evidence of oscillatory corrections,
forming the intermittent succession of bursts of acoustic emission, when
approaching the rupture, has been clearly identified.

In Fig. 2, we present the energy release rate, which is the derivative of our
formula (6) for the cumulative energy release. The result is obained by fitting
to the experimental curve reported at the right bottom in figure 1 of the paper
by Johansen and Sornette (2000). These experimental data explicitly show an
oscillatory behaviour of the energy release rate before the rupture. The best
fit for the five parameters $\alpha$, $\beta$, $\mu$, $\tau$, and $\omega$
have been obtained for 0.7, 1.9, -0.045, 7.5 and 15, respectively, under the
given pressure of rupture at 673 bars. The parameter $\tau$ satisfies Eq. (4).
The upper pressure $p_0\approx 665$ bars, measured before the rupture, corresponds
to the value $y_0\approx 0.0119$, which is sufficiently small, in agreement with
condition (4).

\section{Conclusion}

The parameters can change for different materials as well as for different
experimental setups. However, the three key physical parameters: $\alpha$,
$\omega$, and $\tau$ should be universal, not essentially depending on the
details of the external stress in the critical region. To prove this assumption,
it is necessary to accomplish a set of experiments for appropriate materials.
Selecting particular materials, we should keep in mind that: (i) the width of
the critical region is linearly proportional to the strength of disorder,
(ii) long-range elastic interactions are responsible for log-periodic corrections,
(iii) the more brittle is the fracturing process, the stronger are the elastic
interactions and stronger the energy release fluctuates, (iv) the fluctuations
in the energy release are also stronger when a high value of a material
characteristic is alternated with a low value of that characteristic. Thus,
elastic-porous materials are expected to be the best choice.

\newpage

\begin{center}
{\large{\bf Figure Captions}}
\end{center}

\vskip 2cm

{\bf Fig. 1}. General form of the dimensionless energy release corresponding to
Eq. (6), with $\alpha=0.7$, $\omega=6.9$ and $\tau=6.6$ (continuous line) and
to Eq. (7), with $\alpha=0.7$, $\omega =8$, and $\tau_0=0.8$  (broken line).

\vskip 2cm

{\bf Fig. 2}. Dimensionless energy release rate from Eq. (5), fitted to the
experimental data reported by Johansen and Sornette (2000).

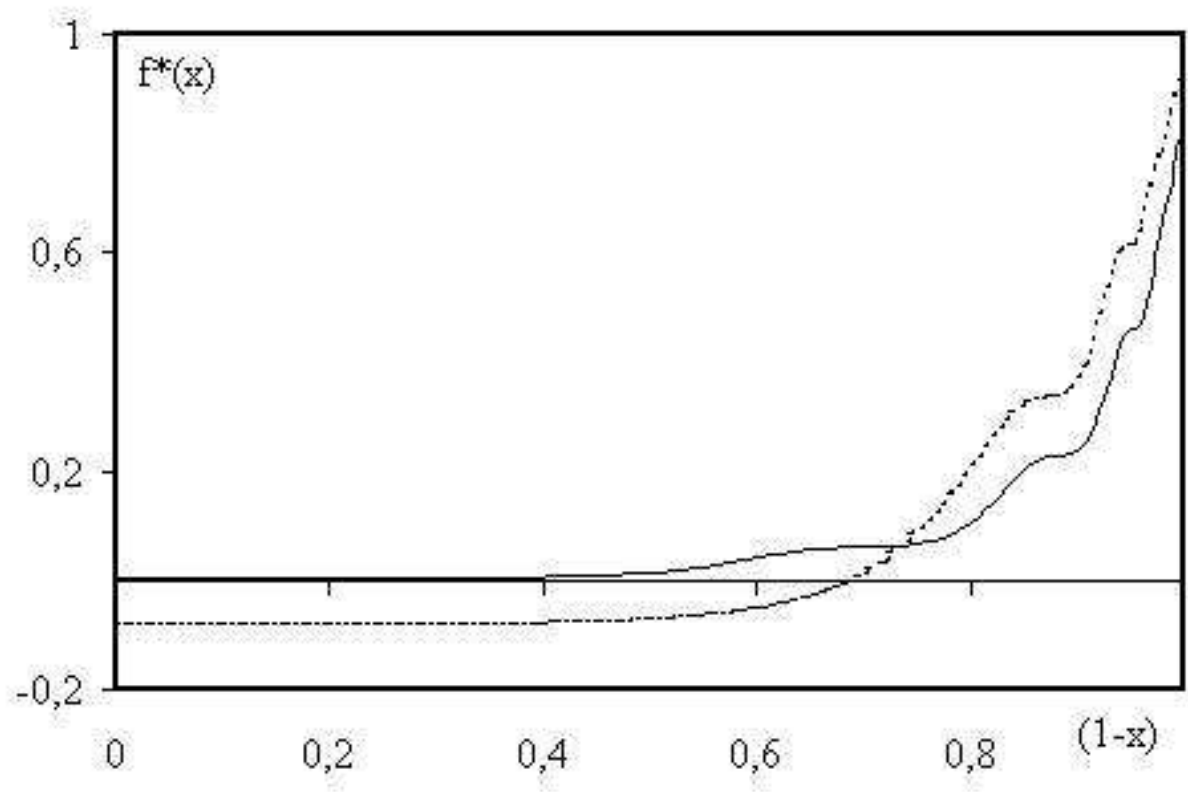
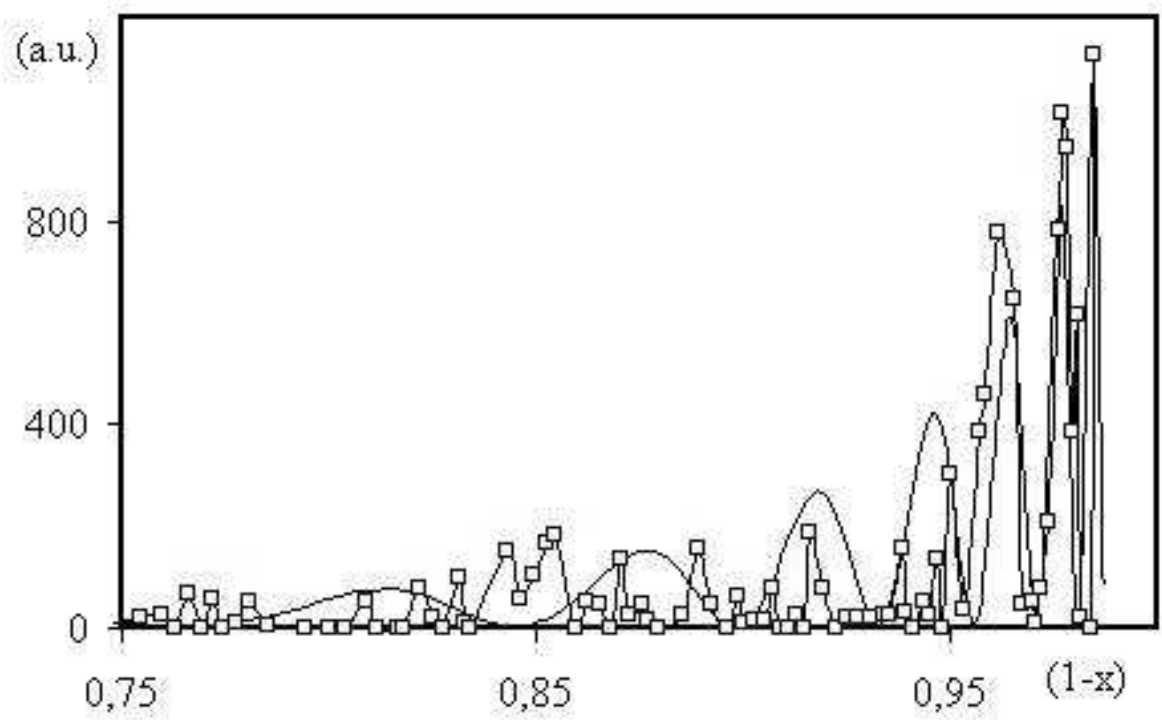
\end{document}